\documentstyle[11pt,newpasp,twoside]{article}
\reversemarginpar

\begin{document}
\title{Type II Quasars from the SDSS}
\author{Nadia L. Zakamska, Michael A. Strauss}
\affil{Princeton University Observatory, Princeton, NJ 08544}
\author{Julian H. Krolik, Timothy M. Heckman}
\affil{Department of Physics and Astronomy, Johns Hopkins University, 3400 North Charles st., Baltimore, MD 21218-2686}

\begin{abstract}
Type II quasars are suggested by unification models of Active Galactic Nuclei as luminous analogs of Seyfert 2 galaxies. A number of different methods have been used hitherto to discover type II quasars, but only a handful have been found. We selected about 150 type II quasar candidates from the SDSS spectroscopic database; these objects have strong narrow high-ionization emission lines. We describe the selection procedure and estimate intrinsic luminosities of these optically obscured AGN. A campaign to perform a multi-wavelength follow-up of the sample is now underway.

\end{abstract}

\section{Introduction}

Unification models of Active Galactic Nuclei (AGN) summarize many of the observed properties of AGN in terms of the intrinsic luminosity and the orientation relative to the line of sight (e.g., Antonucci 1993). In particular, there exists a class of low-luminosity AGN (type 2 Seyfert galaxies) which manifest themselves in the optical as objects with narrow high-ionization emission lines. In the context of unification models it is believed that the line of sight to these objects happens to pass through a lot of obscuring material, so that the ionizing radiation itself and the region that emits broad emission lines are shielded from the observer. If the same unification model applies to high-luminosity AGN (quasars) there should exist high-luminosity obscured AGN (type II quasars), which could account for a large fraction of the hard X-ray background if they exist in significant numbers. 

Campaigns to look for type II quasars at different wavelengths have resulted in discovery of a few tens of candidates (e.g., McCarthy 1993, Kleinmann et al. 1988, Stern et al. 2002, Norman et al. 2002). In all cases multi-wavelength follow-up has been an important part of the confirmation of the candidates. 

We compiled a sample of about 150 type II quasar candidates from the SDSS spectroscopic database in the redshift range $0.3<z<0.8$. We outline the selection process, discuss optical properties of these objects and describe prospects for future work. 

\begin{figure}
\plotfiddle{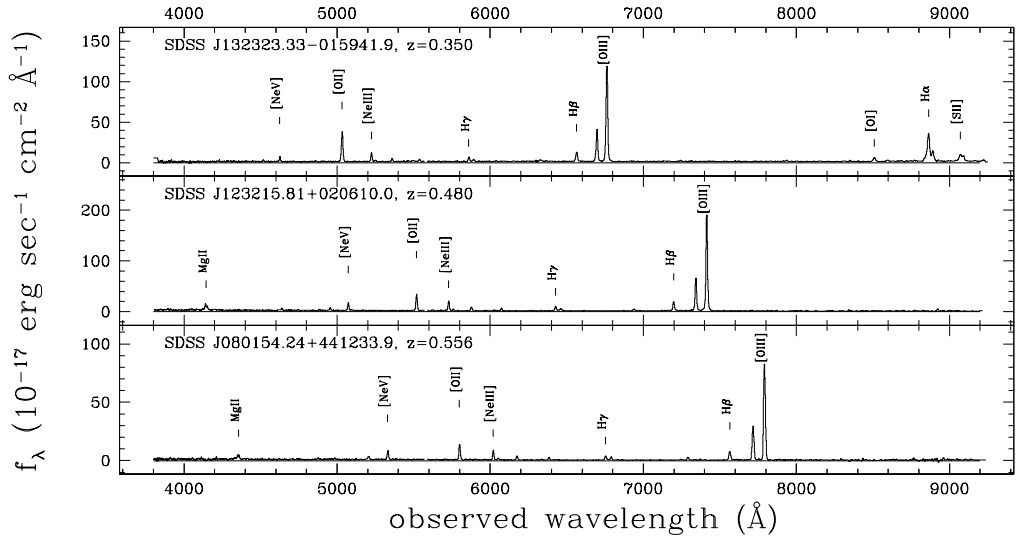}{2.4in}{0}{70}{70}{-205}{-297}
\caption{Example spectra of type II quasar candidates from the SDSS.}
\end{figure}

\section{Optical selection and intrinsic luminosities}

The Sloan Digital Sky Survey (SDSS; York et al. 2000) is an ongoing survey to image 10,000 deg$^2$ of the sky and to obtain spectra for $10^6$ galaxies and $10^5$ quasars. On each 7 deg$^2$ of the sky, about 600 objects are targeted for follow-up spectroscopy based on photometric information. A small number of objects are targeted because they have counterparts in radio or X-ray surveys (Abazajian et al. 2003 and references therein).

We looked for objects with emission line properties similar to those of Seyfert 2 galaxies. In particular, we required that permitted lines do not have underlying broad components. Based on line ratios and presence of high-ionization lines such as [NeV]3426, we distinguished type II AGN from other types of narrow-emission-line objects (star-forming galaxies, narrow-line Seyfert 1 galaxies). We restricted ourselves to the redshift range $0.3<z<0.8$; we thus had spectroscopic coverage of the [OIII]5007 emission line in all spectra to enable statistical studies based on this line. The search resulted in 300 type II AGN candidates among 40,000 SDSS spectra in the same redshift range; example spectra are shown in Figure 1. Although many of these objects have counterparts in the FIRST survey, the radio flux levels are typically a few mJy, and almost all objects in our sample are radio-quiet.

The best way to estimate the intrinsic luminosity of obscured AGN would be to observe them in the mid-IR where the absorbed radiation is reemitted by heated dust particles. Alternatively, one would study type II AGN in hard X-rays which are not affected by obscuration. In the optical, since the narrow emission lines are emitted from the extended (and therefore largely unobscured) region, we can use them as proxies to estimate the intrinsic luminosities of AGN. In Figure 2 we show that for a comparison sample of about 2000 broad-line AGN in the same redshift range ($0.3<z<0.8$) there is a correlation between the [OIII]5007 line luminosity and the broad-band luminosity. Quasars are on the high-luminosity side of this diagram; they are traditionally defined as objects with $M_B<-23$, although the separation between Seyfert 1 galaxies and quasars is somewhat arbitrary. From Figure 2 we find that the corresponding [OIII]5007 luminosity is $3\times 10^8 L_{\odot}$.

\begin{figure}
\plotfiddle{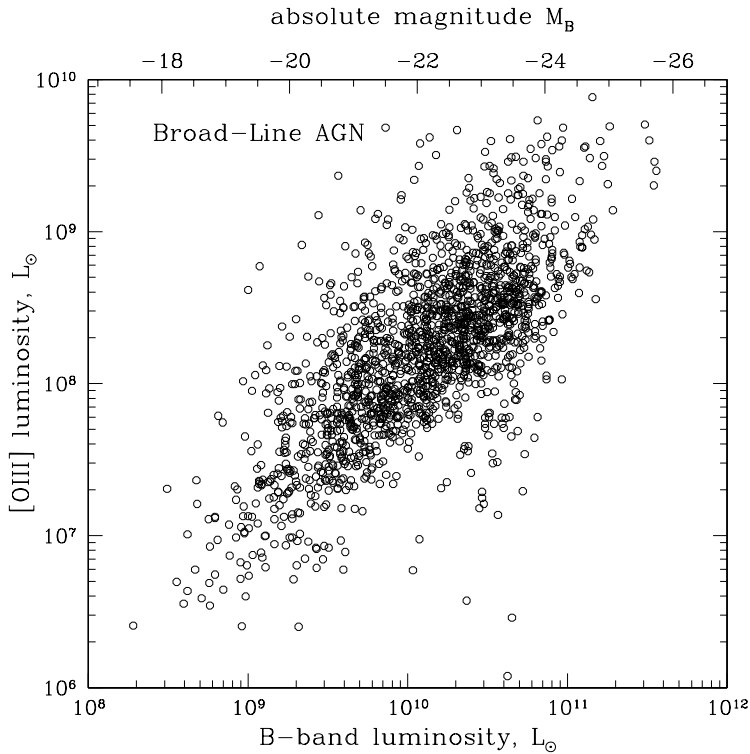}{3.5in}{0}{50}{50}{-160}{-85}
\caption{For broad-line AGN, [OIII]5007 luminosity vs rest-frame B-band luminosity.}
\end{figure}

On Figure 3 we plot the [OIII]5007 luminosity distribution for the 300 type II AGN from our sample and for the comparison sample of broad-line AGN. We find that about 50\% of the objects in our sample have [OIII]5007 line luminosities comparable to those of luminous quasars; thus, about 150 objects in our sample can be classified as type II quasars based on their line properties and intrinsic luminosities. 

Type II quasars are optically faint and thus are difficult to identify in optical surveys (the median magnitude of our sample is $\left<r\right>=20.4$). Selecting such a large number of these objects is possible because of the outstanding size of the spectroscopic survey, part of which is dedicated to the search for unusual objects. Based on the line-luminosity criterion $L$([OIII])$>3\times 10^8 L_{\odot}$ we have identified about 30 type II quasar candidates in the published samples of nearby ($z\la 0.1$) AGN (de Grijp et al. 1992, Whittle 1992), including such well-studied AGN as Mrk34, Mrk463 and Mrk477.

\begin{figure}
\plotfiddle{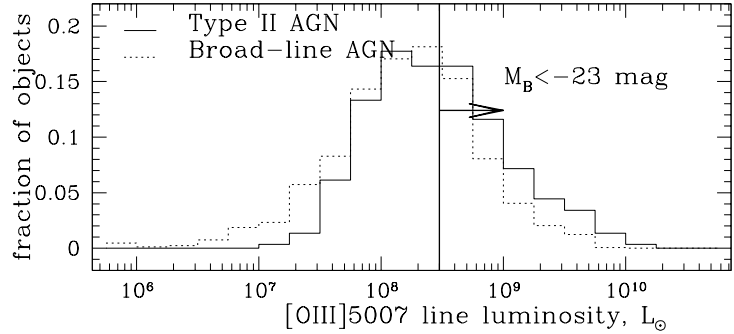}{1.5in}{0}{50}{50}{-160}{-225}
\caption{Distribution of the [OIII]5007 line luminosity for our sample type II AGN and for a comparison sample of broad-line AGN.}
\end{figure}

\section{Conclusions}

In these proceedings we described the selection process that allowed us to select a large number of type II quasar candidates from the SDSS spectroscopic database based on their emission line properties. We then studied their properties in the optical and in particular attempted to estimate the intrinsic luminosity of these objects based on the optical emission lines alone. In Zakamska et al. (2003) we discuss other optical properties of the objects in the sample, such as line ratios, line shapes, continuum properties and the contribution from the host galaxy. We have selected 150 type II quasar candidates up to date and expect to expand the sample to about 450 objects upon completion of the SDSS. 

Our campaign to perform multi-wavelength follow-up of this sample is now underway. We will study properties of these objects at wavelengths ranging from radio and infrared to hard X-rays. Multi-wavelength studies of a large well-defined sample of type II quasars is the best way to probe their physical nature and to make a connection between samples selected by different methods. We will study host galaxies and environments of type II quasars and attempt to identify just which properties of the galactic host and its environment are associated with the AGN activity. Finally, we will use the sample for statistical studies of the AGN population, in particular, we will attempt to constrain the ratio of type I and type II AGN at high luminosities.

Funding for the SDSS is provided by the Alfred P. Sloan Foundation,
NASA, NSF, DoE, Monbukagakusho, the Max Planck Society and the Participating Institutions.  The SDSS web site is http://www.sdss.org/.

\enlargethispage*{0.5in}


\begin{references}
\reference Abazajian, K. et al. 2003, \aj, in press 
\reference Antonucci, R. 1993, \araa,31,473
\reference de Grijp, M.H.K., Keel, W.C., Miley, G.K., et al. 1992, A\&ASS,96,389
\reference Kleinmann, S.G., Hamilton, D., Keel, W.C., et al. 1988,\apj,328,61
\reference McCarthy, P.J. 1993, \araa,31,639
\reference Norman, C., Hasinger, G., Giacconi, R., et al. 2002,\apj,571,218
\reference Stern, D., Moran, E.C., Coil, A.L., et al. 2002,\apj,568,71
\reference Whittle, M. 1992, \apjs,79,49
\reference Zakamska, N.L., Strauss, M.A., Krolik, J.H., et al. 2003, \aj, in press, \\ astro-ph/0309551
\end{references}
\end{document}